\documentclass[a4paper,fleqn,usenatbib]{mnras}

\usepackage{newtxtext,newtxmath}

\usepackage[T1]{fontenc}
\usepackage{ae,aecompl}

\usepackage{graphicx}	
\usepackage{amsmath}	
\usepackage{amssymb}	

\title[PWNe inside SNRs as CR PeVatrons]{Pulsar Wind Nebulae inside Supernova Remnants as Cosmic-Ray PeVatrons}

\author[Y. Ohira et al.]{
Yutaka Ohira$^{1,2}$,\thanks{E-mail: y.ohira@eps.s.u-tokyo.ac.jp (YO)}
Shota Kisaka$^{2}$,
and Ryo Yamazaki$^{2}$
\\
$^{1}$Department of Earth and Planetary Science, The University of Tokyo, 7-3-1 Hongo, Bunkyo-ku, Tokyo 113-0033, Japan\\
$^{2}$Department of Physics and Mathematics, Aoyama Gakuin University, 5-10-1 Fuchinobe, Sagamihara 252-5258, Japan
}

\date{}

\begin{document}
\label{firstpage}
\pagerange{\pageref{firstpage}--\pageref{lastpage}}
\maketitle

\begin{abstract}
We propose that cosmic-ray PeVatrons are pulsar wind nebulae (PWNe) inside supernova remnants (SNRs).
The PWN initially expands into the freely expanding stellar ejecta. 
Then, the PWN catches up with the shocked region of the SNR, where particles can be slightly accelerated by the back and forth motion between the PWN and the SNR, and some particles diffuse into the PWN. 
Afterwards the PWN is compressed by the SNR, where the particles in the PWN are accelerated by the adiabatic compression.  
Using a Monte Carlo simulation, we show that particles accelerated by the SNR to $0.1~{\rm PeV}$ can be reaccelerated to $1~{\rm PeV}$ until the end of the PWN compression. 
\end{abstract}




\section{Introduction}

The origin of cosmic-ray (CR) PeVatrons is a long standing problem in the astrophysics. 
The CR spectrum has a spectral break at $\sim 10^{15}$\,eV $=1$\,PeV (so called, the knee energy). 
The diffusive shock acceleration (DSA) at supernova remnants (SNRs) is believed to be the acceleration mechanism of CRs up to the knee energy \citep{1977ICRC...11..132A,1977DoSSR.234.1306K,1978MNRAS.182..147B,1978ApJ...221L..29B}. 
Although recent gamma-ray observations support the idea \citep{2011MNRAS.410.1577O,2013Sci...339..807A}, there are still many problems. 
One is the knee problem. 
It was estimated that SNRs cannot accelerate CRs to the knee energy for a parallel shock without strong magnetic field  \citep{1983A&A...125..249L}. 
In order to accelerate CRs to the knee energy, magnetic fields must be amplified in the shock upstream region. 
Several mechanisms of the magnetic field amplification in the shock upstream region have been proposed \citep{2004MNRAS.353..550B,2010PPCF...52l4006M,2010ApJ...721L..43O,2012ApJ...758...97O}. 
However, no simulations have demonstrated that the upstream magnetic field is sufficiently amplified to accelerate CRs to the knee energy. 
In contrast to the shock upstream region, magnetic fields are expected to be easily amplified to the equipartition level in the shock downstream region \citep{2007ApJ...663L..41G,2009ApJ...695..825I,2012ApJ...747...98G,2013ApJ...765L..20C,2016ApJ...817..137O}. 
Super-Alfv{\'e}nic turbulence amplifies the field by stretching the field line. 
The downstream turbulence is generated by interactions between upstream density fluctuations and the shock front. 
In addition, the downstream turbulence is generated by the Rayleigh-Taylor instability at the contact discontinuity.  
As a result, the magnetic field in the shocked region is amplified by the turbulence. 
If only the downstream magnetic field is amplified, the acceleration time scale of DSA is predominantly determined by the upstream residence time of accelerated particles, which depends on the shock velocity, $u_{\rm sh}$, and the diffusion coefficient, $D$,  \citep{2017JHEAp..13...17O,1983RPPh...46..973D}. 
Then the acceleration time scale is given by
\begin{eqnarray}
t_{\rm acc} &\approx& \frac{4D}{u_{\rm sh}^2}= \frac{4cE}{3eu_{\rm sh}^2B_{\rm up}} \nonumber \\
&\approx& 10^4\,{\rm yr} \left(\frac{E}{1\,{\rm PeV}}\right)\left(\frac{u_{\rm sh}}{3\times 10^3\,{\rm km}\,{\rm s}^{-1}}\right)^{-2}\left(\frac{B_{\rm up}}{3\,\mu{\rm G}}\right)^{-1}~~,
\end{eqnarray}
where we assume the shock compression ratio of 4, the Bohm diffusion coefficient, $D_{\rm B}=cE/3eB_{\rm up}$, and $c, e, E$ and $B_{\rm up}$ are the speed of light, elementary charge, particle energy, and upstream magnetic field strength, respectively. 
After the free expansion phase ($t>t_{\rm S}$), the velocity of the forward shock decreases with time.  
The Sedov time, $t_{\rm S}$, is given by
\begin{equation}
t_{\rm S} \approx 10^3\,{\rm yr} \left(\frac{E_{\rm SN}}{10^{51}\,{\rm erg}}\right)^{-\frac{1}{2}} \left(\frac{M_{\rm ej}}{3\,{\rm M}_{\sun}}\right)^{\frac{5}{6}} \left(\frac{n}{0.1\,{\rm cm}^{-3}}\right)^{-\frac{1}{3}}~~,
\end{equation}
where $E_{\rm SN}, M_{\rm ej}$ and $n$ are the explosion energy, ejecta mass, and the ambient number density, respectively \citep{1995PhR...256..157M}. 
From the condition, $t_{\rm acc}=t_{\rm S}$, the maximum energy of particles accelerated at the forward shock is given by
\begin{equation}
E_{\rm max} \approx 0.1\,{\rm PeV}\left(\frac{E_{\rm SN}}{10^{51}\,{\rm erg}}\right)^{\frac{1}{2}} \left(\frac{M_{\rm ej}}{3\,{\rm M}_{\sun}}\right)^{-\frac{1}{6}} \left(\frac{n}{0.1\,{\rm cm}^{-3}}\right)^{-\frac{1}{3}} \left(\frac{B_{\rm up}}{3\,\mu{\rm G}}\right)~~,
\label{eq:emax1}
\end{equation}

where $u_{\rm sh}=(10E_{\rm SN}/3M_{\rm ej})^{1/2}$ is assumed \citep{1995PhR...256..157M}. 
This is about 10 times smaller than the knee energy. 
The maximum energy weakly depends on parameters of supernovae 
and their environment except for the upstream magnetic field. 
Therefore, in oder for SNRs to accelerate CRs to the knee energy, 
the upstream magnetic field needs to be amplified to about $10^2\,\mu$G.

The other possible solution for the knee problem is DSA at the perpendicular shocks \citep{1987ApJ...313..842J}. 
Since accelerated particles cannot propagate to the far upstream region, 
the acceleration time scale becomes small for the perpendicular shock.
Although the injection to DSA was thought to be difficult for the perpendicular shock, 
it was shown by three-dimensional hybrid simulations that 
particles are injected to DSA at the perpendicular shock in a partially ionized plasma, so that 
particles are rapidly accelerated there \citep{2016ApJ...827...36O}. 
However, DSA at the perpendicular shocks has another problem. 
For such cases, the maximum energy is limited by the size of acceleration region, $R$. 
The available potential drop is $\Delta \phi=RB_{\rm up}u_{\rm sh}/c$, 
so that the maximum energy of accelerated protons is given by 
\begin{eqnarray}
E_{\rm max} &=& e\Delta \phi \nonumber \\
&\approx& 0.1\,{\rm PeV} \left(\frac{R}{10\,{\rm pc}}\right) \left(\frac{B_{\rm up}}{3\,\mu {\rm G}}\right)\left(\frac{u_{\rm sh}}{3\times 10^3\,{\rm km}\,{\rm s}^{-1}}\right)~~, 
\label{eq:emax2}
\end{eqnarray}
which is again 10 times smaller than the knee energy. 
Since $u_{\rm sh}=(10E_{\rm SN}/3M_{\rm ej})^{1/2}$ and $R=u_{\rm sh} t_{\rm S}$, equation (\ref{eq:emax2}) becomes identical to equation (\ref{eq:emax1}). 
In order to accelerate CRs to the knee energy at the perpendicular shock, we need an exceptional condition \citep{2015ApJ...809...29T}.

In this paper, we propose a reacceleration mechanism from $0.1$\,PeV to $1$\,PeV by pulsar wind nebulae (PWNe) inside SNRs. 
Recent observations of young $\gamma$-ray pulsars suggest that most core-collapse supernovae generate pulsars and their spindown luminosity is typically $L_{\rm sd}\sim 3\times 10^{38}$\,erg\,s$^{-1}$ \citep{2011ApJ...727..123W}. 
As mentioned above, the magnetic field in the shocked region of SNRs is strong enough to scatter high-energy particles. 
The magnetic field in young PWNe is also strong compared with that in the interstellar medium, 
which is about $B_{\rm PWN}\sim 10^2\,{\rm \mu G}$ \citep{2010ApJ...715.1248T,2014JHEAp...1...31T}. 
The PWN initially expands into the freely expanding stellar ejecta toward the shocked region of the SNR. 
Since this system can be interpreted as two walls approaching each other, 
particles are accelerated, shuttling between the PWN and the shocked region of the SNR. 
After the PWN reaches the reverse shock of the SNR, the PWN is compressed and particles inside the PWN are accelerated by the adiabatic compression \citep{2001ApJ...563..806B,2001A&A...380..309V}. 
In the next section, using Monte Carlo simulation, we show that the PWN-SNR system actually accelerates particles 
from $0.1$\,PeV to $1$\,PeV.

\section{Monte Carlo simulations} 
In order to investigate the particle acceleration by the PWN-SNR system, 
we first provide evolution of an SNR and a PWN inside the SNR. 
As a first step, we consider a spherically symmetric structure. 
For constant ejecta and ambient density profiles, the approximate time evolution of the forward and reverse shock radii, $R_{\rm SNR,fs}$ and $R_{\rm SNR,rs}$, are given by \citet{1995PhR...256..157M}. 
They got
\begin{eqnarray}
\frac{R_{\rm SNR,fs}}{R_{\rm S}} &=&1.37\,\frac{t}{t_{\rm S}} \left\{1+0.60\,\left(\frac{t}{t_{\rm S}}\right)^{3/2} \right\}^{-2/3} ~~,\label{rsnrfs}\\
\frac{R_{\rm SNR,rs}}{R_{\rm S}} &=&1.24\,\frac{t}{t_{\rm S}}\left\{1+1.13\,\left(\frac{t}{t_{\rm S}}\right)^{3/2} \right\}^{-2/3} ~~,
\label{eq:ed}
\end{eqnarray}
for the free expansion phase $(t < t_{\rm S})$, and 
\begin{eqnarray}
\frac{R_{\rm SNR,fs}}{R_{\rm S}} &=& \left(1.56\,\frac{t}{t_{\rm S}}-0.56 \right)^{2/5} ~~,\\
\frac{R_{\rm SNR,rs}}{R_{\rm S}} &=&\frac{t}{t_{\rm S}} \left\{ 0.78 - 0.03\,\frac{t}{t_{\rm S}} - 0.37\,\ln{ \left( \frac{t}{t_{\rm S}} \right) } \right \} ~~,
\end{eqnarray}
for the Sedov phase$(t \geq t_{\rm S})$, where $R_{\rm S}$ is given by \citep{1995PhR...256..157M}, 
\begin{eqnarray}
R_{\rm S}&=&0.805\,t_{\rm S}\left(\frac{10E_{\rm SN}}{3M_{\rm ej}}\right)^{1/2} \nonumber \\ 
&\approx& 7\,{\rm pc} \left(\frac{M_{\rm ej}}{3M_{\sun}}\right)^{1/3}\left(\frac{n}{0.1{\rm cm}^{-3}}\right)^{-1/3}~~.
\end{eqnarray}
For a constant spindown luminosity of a pulsar and an uniform ejecta profile, 
the analytical solution for the time evolution of the PWN radius, $R_{\rm PWN}$, is given by \citep{2001A&A...380..309V}, 
\begin{equation}
\frac{R_{\rm PWN}}{R_{\rm S}} = 1.04\,\left(\frac{L_{\rm sd}t_{\rm S}}{E_{\rm SN}}\right)^{1/5}  \left(\frac{t}{t_S}\right)^{6/5}~~,
\label{rpwn}
\end{equation}
where the solution can be applied until the PWN interacts with the reverse shock of the SNR.
It should be noted that the PWN radius, $R_{\rm PWN}$, does not significantly depend on the spindown luminosity, $L_{\rm sd}$.  

\begin{figure}
\centering
\includegraphics[width=\columnwidth]{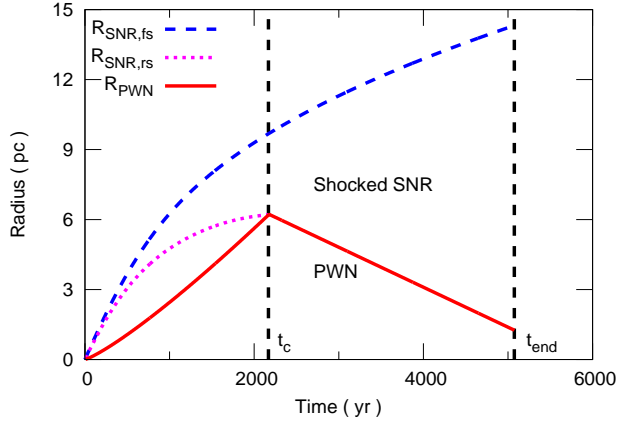}
\caption{Time evolution of the PWN radius (red solid) with a spindown luminosity of $L_{\rm sd}=3\times10^{38}$\,erg\,s$^{-1}$, 
and radii of the forward (blue dashed) and the reverse shocks (magenta dotted) of the SNR with an explosion energy of $E_{\rm SN} = 10^{51}$\,erg,  an ejecta mass of $M_{\rm ej}=3\,M_{\sun}$, and an ambient number density of $n=0.1$\,cm$^{-3}$. }
\label{fig:1}
\end{figure}

Fig.~\ref{fig:1} shows the time evolutions of the forward and reverse shock radii of the SNR and the radius of PWN, 
where we assume the uniform ejecta profile with the ejecta mass of $M_{\rm ej}=3\,{\rm M}_{\sun}$, the explosion energy of $E_{\rm SN}=10^{51}$\,erg, the uniform ambient matter profile with the density of $n=0.1~{\rm cm}^{-3}$, and the constant pulsar spindown luminosity of $L_{\rm sd}=3\times 10^{38}$\,erg\,s$^{-1}$. 
For these parameters, the PWN catches up with the reverse shock of the SNR at $t_{\rm c} \approx 2\times 10^3$\,yr. 
Afterwards, the PWN is compressed by the larger pressure of the shocked region of the SNR. 
In this paper, the constant spindown luminosity is assumed for $t \leq t_{\rm c}$. 
Then we simply assume that the velocity of the PWN during the compression is $v_{\rm PWN} = -v_{\rm PWN}(t_{\rm c})/2$ and the final size of the PWN is $R_{\rm PWN}(t_{\rm end}) = R_{\rm PWN}(t_{\rm c}) / 5$. 
Then, the PWN size becomes $\approx 1.2$\,pc at $t_{\rm end} \approx 5\times 10^3$\,yr. 
These assumptions are reasonable to simulate evolution of a spherical PWN \citep[e.g.][]{2009ApJ...703.2051G}. 

\begin{figure}
\centering
\includegraphics[width=\columnwidth]{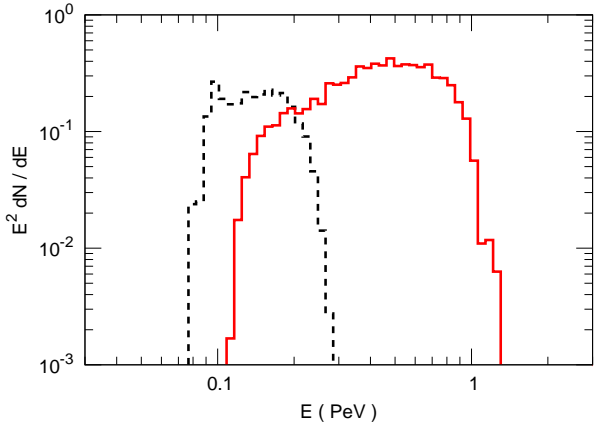}
\caption{Energy spectra of reaccelerated particles for Model A. 
The black dashed and red solid histograms are energy spectra at $t=t_{\rm c}$ and $t_{\rm end}$, respectively. 
The initial energy is $0.1$\,PeV.}
\label{fig:2}
\end{figure}

We next give the velocity field in the PWN-SNR system.
Before the PWN reaches the reverse shock of the SNR, the expansion velocity of the shocked ejecta just outside the reverse shock in the observer frame is given by 
\begin{equation}
v_{\rm SNR,shocked} = \frac{R_{\rm SNR,rs}/t-v_{\rm SNR,rs}}{4} + v_{\rm SNR,rs}~~, 
\end{equation}
where we assume the compression ratio at the SNR reverse shock is $4$, and $v_{\rm SNR,rs}=dR_{\rm SNR,rs}/dt$ is the propagation velocity of the reverse shock in the observer frame. 
In this paper, the velocity field between the forward and reverse shocks is approximated, for simplicity, as a linear interpolation between $v_{\rm SNR, shocked}$ at $r=R_{\rm SNR,rs}+0$ and $3v_{\rm SNR,fs}/4$ at $r=R_{\rm SNR,fs}-0$. 
The expansion velocity of the PWN is given by $v_{\rm PWN}=dR_{\rm PWN}/dt$ and the velocity field in the PWN is assumed to be uniform. 

After the PWN interacts with the reverse shock of the SNR, the velocity between the forward shock and the PWN is approximately given by the linear interpolation between $v_{\rm PWN} = -v_{\rm PWN}(t_{\rm c})/2$ and $3v_{\rm SNR,fs}/4$. 
The velocity field in the PWN is assumed as
\begin{equation}
{\vec v}_{\rm PWN,in}(t,{\vec r}) = -\frac{v_{\rm PWN}(t_{\rm c})}{2}\frac{{\vec r}}{R_{\rm PWN}(t)}~~.
\end{equation}

Since the shocked region of the SNR and the PWN region are expected to be highly turbulent \citep{2014MNRAS.438..278P}, 
motion of high-energy particles could be approximated as the random walk. 
Using the above hydrodynamical structure, we perform a test-particle Monte Carlo simulation. 
Simulation particles are isotropically scattered in the local fluid frame. 
The scattering time is assumed to be the Bohm scattering, $t_{\rm sc} = \Omega_{\rm c}^{-1} (E/m_{\rm p} c^2)$, 
where $\Omega_{\rm c}\approx 10^{-2}\,{\rm s}^{-1} (B/1\,\mu{\rm G})$ is the proton cyclotron frequency, $E$ is the particle energy, and $m_{\rm p}$ is the proton mass. 
{
Once particles escape from the SNR, we do not follow the particles. 
However, no particles escape from the SNR in this paper. 
}

\begin{figure}
\centering
\includegraphics[width=\columnwidth]{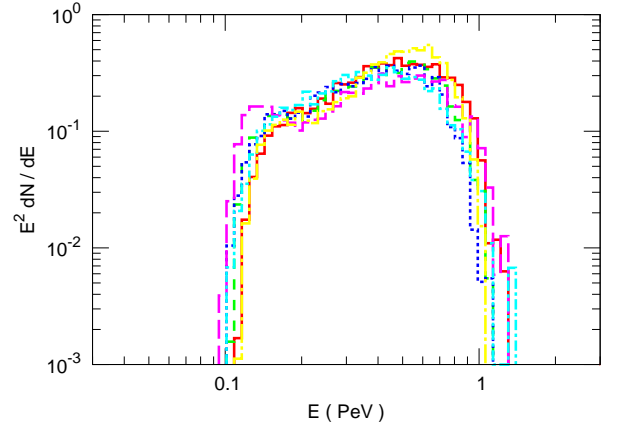}
\caption{Energy spectra of reaccelerated particles at the end of the PWN compression, $t=t_{\rm end}$. 
The red solid, green dashed, blue dotted, magenta long-dashed, cyan dot-dashed, and yellow dot-long-dashed histograms are for Model A, B, C, D, E, and F. 
The red histogram in this figure is the same as in Fig.~\ref{fig:2}.
The parameters for each model are tableted in Table~\ref{tab:1}. 
}
\label{fig:3}
\end{figure}
\begin{table}
\centering
\begin{tabular}{|c|c|c|c|c|c|c|}
\hline Model & $M_{\rm ej}$ & $n$ & $L_{\rm sd}$ & $t_{\rm inj}$  & $B_{\rm SNR}$ & $B_{\rm PWN}$ \\
 & ($M_{\sun}$) & (${\rm cm}^{-3}$) &  (erg\,s$^{-1}$) &  (s) &  ($\mu$G)&  ($\mu$G)\\
\hline A & 3 & 0.1 & $3\times{10}^{38}$ & $5\times10^{10}$ & $300$&$150$\\
\hline B & 10 & 0.1 & $3\times{10}^{38}$ & $1\times10^{11}$ & $300$&$150$\\
\hline C & 3 & 1.0 & $3\times{10}^{38}$ & $2.5\times10^{10}$ & $300$&$150$\\
\hline D & 3 & 0.1 & $3\times{10}^{37}$ & $5\times10^{10}$ & $300$&$150$\\
\hline E & 3 & 0.1 & $3\times{10}^{38}$ & $5\times10^{10}$ & $150$&$150$\\
\hline F & 3 & 0.1 & $3\times{10}^{38}$ & $5\times10^{10}$ & $300$&$50$\\

\hline
\end{tabular}
\caption{List of parameters for Fig. 3.}
\label{tab:1}
\end{table}

In this paper, we set the magnetic field to be $B_{\rm SNR}=3\times10^2\,\mu$G and $B_{\rm PWN}=1.5\times10^2\,\mu$G in the shocked region of the SNR and PWN, respectively, 
and $B_{\rm ej}=0$ in the freely expanding ejecta. 
Hence, particles are not scattered in the freely expanding ejecta which disappears after the PWN interacts with the reverse shock of the SNR. 
Since the forward shock of the SNR can accelerate particles to about $0.1$\,PeV (see introduction), we set the initial energy to be $0.1$\,PeV. 
The accelerated particles are advected toward the downstream region of the forward shock. 
Furthermore, they are expected to be advected (or diffuse) to the reverse shocked region because of the Rayleigh-Taylor instability and turbulence. 
In the next section, we will discuss on the amount of particles which the turbulence carries from the forward shock to the revere shock front. 
In this paper, we impulsively inject simulation particles isotropically on the reverse shock sphere, $r = R_{\rm SNR,rs}$ at $t_{\rm inj}=1.67\times10^3$\,yr instead of solving the particle transport from the forward shock to the reverse shock, 
which should be addressed in the future. 
About half of the injected particles initially diffuse to the reverse shocked region, and the rest of particles run to the freely expanding ejecta.  


Fig.~\ref{fig:2} shows energy spectra of the accelerated particles for Model A, parameters of which are listed in Table~\ref{tab:1}. 
The black dashed histogram shows the energy spectrum at the time when the PWN reaches the reverse shock of the SNR ($t=t_{\rm c}$). 
They are accelerated up to twice the initial energy by the back and forth motion between the PWN and the shocked region of the SNR. 
The energy gain in each cycle is $\Delta E/E \sim \Delta v/c$, where $\Delta v=v_{\rm SNR,shocked}-v_{\rm PWN}$ is the relative velocity. 
The time scale in each cycle is $\Delta t \sim \Delta R/c$, where $\Delta R=R_{\rm SNR,rs}-R_{\rm PWN}$ is the relative distance. 
Then, the acceleration time scale for the reciprocation is given by $t_{\rm acc}=\Delta t (E/\Delta E) \sim \Delta R/ \Delta v$ that is the same as the dynamical time scale in which the PWN catches up with the SNR reverse shock. 
Since the acceleration time scale, $t_{\rm acc}$, does not significantly depend on the magnetic field strength and the particle energy as long as the scattering time is smaller than the time scale of the reciprocation between the SNR and the PWN, $\Delta t$, it takes $t_{\rm acc}$ to accelerate particles to twice the initial energy. 
Therefore, the maximum energy during the approaching phase becomes twice the initial energy, 
which does not significantly depend on parameters of the PWN-SNR system.

The red solid histogram in Fig.~\ref{fig:2} shows the energy spectrum at $t=t_{\rm end}$. 
They are further accelerated to $1$\,PeV by the PWN compression. 
The particle energy is increased by a factor $R_{\rm PWN}(t_{\rm c})/R_{\rm PWN}(t_{\rm end})=5$ during the compression,  
so that the particles are finally accelerated to ten times the initial energy. 
Hence, the maximum energy of particles accelerated by the PWN-SNR system is given by 
\begin{equation}
E_{\rm max} \sim 1\,{\rm PeV} \left(\frac{E_{\rm inj}}{0.1\,{\rm PeV}} \right) \left(\frac{ R_{\rm PWN}(t_{\rm c})/R_{\rm PWN}(t_{\rm end}) }{5} \right)~~, 
\end{equation}
where $E_{\rm inj}$ is the initial energy of particles injected to the PWN-SNR system, 
that corresponds to the maximum energy of particles accelerated by the SNR shocks. 
The compression factor, $R_{\rm PWN}(t_{\rm c})/R_{\rm PWN}(t_{\rm end})$, is determined by the pressure balance between the PWN and the shocked region of the SNR at $t_{\rm end}$. 
The rotational energy of a pulsar, $E_{\rm rot}$, is initially stored in the PWN, 
but the PWN eventually loses its energy by synchrotron radiation. 
The synchrotron cooling time is given by 
\begin{equation}
t_{\rm cool} \approx 1.2\, {\rm kyr} \left(\frac{B}{10^2\, {\rm \mu G}}\right)^{-2} \left(\frac{E}{1\,{\rm TeV}}\right)^{-1} ~~,
\end{equation}
which is smaller than $t_{\rm end}$. 
The characteristic energy is typically a few $10^2\,{\rm GeV}-1\,{\rm TeV}$ \citep{2014JHEAp...1...31T}.
Therefore, most of the rotational energy converts into synchrotron photons during the compression phase of PWNe. 
Then, the remaining energy in the PWN is $\eta E_{\rm rot}$ at $t_{\rm end}$, 
where $\eta$ is the remaining fraction of the order of 0.1 \citep{2009ApJ...703.2051G}. 
From the equation $E_{\rm SN}/R_{\rm SNR,fs}(t_{\rm end})^3 = \eta E_{\rm rot}/R_{\rm PWN}(t_{\rm end})^3$, 
the radius of the PWN at the end of the compression is given by 
\begin{equation}
R_{\rm PWN}(t_{\rm end})= 0.1\,R_{\rm SNR,fs}(t_{\rm end})\left(\frac{\eta E_{\rm rot}/E_{\rm SN}}{10^{-3}} \right)^{1/3} ~~,
\end{equation}
where we set $E_{\rm SN}=10^{51}$\,erg, $E_{\rm rot}=10^{49}$\,erg and $\eta =0.1$. 
Since $R_{\rm SNR,fs}(t_{\rm end})\approx 13\,{\rm pc}$ and $R_{\rm PWN}(t_{\rm c})\approx 6\,{\rm pc}$ in this paper (see Fig.~\ref{fig:1}), the compression factor, $R_{\rm PWN}(t_{\rm c})/R_{\rm PWN}(t_{\rm end})= 5$, is a reasonable approximation. 
If the initial rotational energy of the pulsar is smaller, 
the PWN is more compressed, so that particles are accelerated to higher energies by the compression. 
However, it should be noted that the maximum energy cannot exceed the limitation by the PWN size. 
Once the gyroradius of accelerated particles becomes comparable to the PWN size, 
they start to escape from the PWN.
 
In order to explore the parameter dependence of the above results, we perform other simulations 
with different parameters. 
The parameter sets are tableted in Table~\ref{tab:1}. 
As can be seen in Fig.~\ref{fig:3}, the results do not change significantly as long as 
$R_{\rm PWN}(t_{\rm c})/R_{\rm PWN}(t_{\rm end})=5$ is fixed. 
This is because dynamics of SNR and PWN do not significantly depend on the spindown luminosity 
and the ambient number density (see equations (\ref{rsnrfs})--(\ref{rpwn})). 
Although the ejecta mass dependency of the Sedov time ($t_{\rm S}\propto M_{\rm ej}^{5/6}$) is 
comparatively strong compared with other parameter dependences, the ejecta mass is expected not to be 
distributed widely for core collapse supernovae that leave a neutron star. 
The acceleration by the PWN compression does not depend on the magnetic field strength. 
For $t_{\rm inj} < t < t_{\rm c}$, particles diffuse into the PWN. 
The diffusion length scale is given by $R_{\rm diff}=\sqrt{4D( t_{\rm c} - t_{\rm inj}) }$, 
where $D$ is the diffusion coefficient.  
On the other hand, during the PWN compression phase ($t_{\rm c} < t < t_{\rm end}$), 
the particles escape from the PWN by diffusion. 
The escape time scale, $t_{\rm esc}$ is given by $t_{\rm esc}=R_{\rm diff}^2/4D = t_{\rm c} - t_{\rm inj}$, 
which is independent on the diffusion coefficient and magnetic field strength, 
so that the final spectrum does not depend on the magnetic field strength. 
Therefore, many PWN-SNR systems can be expected to accelerate particles to the knee energy.

\section{Injection at the reverse shock}
In the previous section, we assumed that $0.1$ PeV CRs are injected at the reverse shock of the SNR to reaccelerate them to $1$ PeV. 
In this section, we discuss some injection mechanisms at the reverse shock of the SNR. 
There are four shocks in the PWN-SNR system before the PWN interacts with the reverse shock of the SNR, 
the reverse and forward shocks of the SNR, the termination shock of the pulsar wind, and the forward shock driven by the PWN. 
The termination shock of the pulsar wind can accelerate electrons and positrons in the standard picture, 
but it has not been understood whether protons and heavy nuclei are accelerated by the termination shock or not. 

The reverse shock of the SNR and the forward shock driven by the PWN propagate into the freely expanding ejecta, 
where the magnetic field strength in the shock upstream region is expected to be very weak. 
In this case, we naively expect a weak CR acceleration by the shocks propagating into the freely expanding ejecta. 
However, if the magnetic field is sufficiently amplified by some mechanisms, the reverse shock of the SNR and the forward 
shock driven by the PWN can accelerate CRs to $0.1$ PeV. 
Then, they are further accelerated to PeV CRs by the PWN-SNR system as shown in the previous section. 

The forward shock of the SNR can easily accelerate CRs to $0.1$~PeV as estimated in the introduction. 
They are transported to the reverse shock of the SNR as described in the following. 
Since the magnetic field in the downstream region of the SNR is amplified by turbulence, 
the diffusion coefficient due to the particle diffusion would be in the Bohm limit. 
Then, the diffusion length scale is given by
\begin{equation}
\begin{aligned}
R_{\rm diff,p}  &= \sqrt{4D_{\rm B}t} \\
                       &=2 \times 10^{18}\,{\rm cm} \left(\frac{E}{0.1\,{\rm PeV}}\right)^{\frac{1}{2}} \left(\frac{B}{100\,{\rm \mu G}}\right)^{-\frac{1}{2}} \left(\frac{t}{1\,{\rm kyr}}\right)^{\frac{1}{2}}~~, 
\end{aligned}
\end{equation}
which is comparable to, but still smaller than the distance between the forward and reverse shocks, $\Delta R = R_{\rm SNR,fs}-R_{\rm SNR,rs}\sim$ a few parsecs. 
Here, $D_{\rm B}$ is the Bohm diffusion coefficient of the particle diffusion. 
To be reaccelerated by the back and forth motion between the PWN and the reverse shock of the SNR, 
particles have to be injected within a distance of $l_{\rm p}$ from the reverse shock of the SNR, 
where the diffusion length scale, $l_{\rm p}$, is given by 
\begin{equation}
\begin{aligned}
l_{\rm p} &=\frac{D_{\rm B}}{u_2} \\
                      &= 3.3 \times 10^{17}\,{\rm cm} \left(\frac{E}{0.1\,{\rm PeV}}\right) \left(\frac{B}{100\,{\rm \mu G}}\right)^{-1} \left(\frac{u_2}{10^3\,{\rm km\,s}^{-1}}\right)~~, 
\end{aligned}
\end{equation}
and $u_2$ is the downstream flow velocity in the reverse shock rest frame. 
If the diffusion region, $l_{\rm p}$, overlaps with $R_{\rm diff,p}$ ($l_{\rm p}+R_{\rm diff,p}>\Delta R$), 
CRs accelerated at the forward shock can be transported to the reverse shock. 
Since this condition is not satisfied as long as the magnetic field is strongly amplified, 
particles cannot diffuse to the freely expanding ejecta from the forward shock by the Bohm diffusion. 
If the amplified magnetic field sufficiently decays in the shock downstream region, 
most of particles accelerated at the forward shock can diffuse to the freely expanding ejecta, 
so that they are reaccelerated by the PWN-SNR system. 

Even if the amplified magnetic field does not decay sufficiently, 
the particles accelerated at the forward shock would be transported to the reverse shock by the turbulent diffusion. 
There are two types of diffusion in the downstream region of the SNR, 
the particle diffusion and the turbulent diffusion. 
The particle diffusion due to magnetic turbulence can move particles to other fluid elements, 
so that particles can move from the downstream region to the upstream region.  
On the other hand, in the context of the turbulent diffusion, particles move with a fluid element, 
so that particles cannot penetrate the shock front. 
If there are magnetic turbulence and large-scale fluid turbulence in the downstream region, 
and if the diffusion coefficient of the particle diffusion is smaller than that of the turbulent diffusion, 
the particle diffusion and the turbulent diffusion coexist. 
In a timescale smaller than the eddy turnover time, 
motion of particles can be described by the particle diffusion, 
but it can be described by the turbulent diffusion in a timescale larger than the eddy turnover time. 
If the eddy size of turbulence is about $\Delta R$, 
particles accelerated at the forward shock can move to the vicinity of the revere shock. 
In this case, the injection fraction of particles at the reverse shock is about $l_{\rm p}/\Delta R\sim 0.17$. 
If there is turbulence with an eddy size of $L_{\rm edd}$ in the vicinity of the reverse shock, 
particles within a distance of $l_{\rm turb}=D_{\rm turb}/u_2$ from the reverse shock can diffuse to the reverse shock by turbulence, 
where $D_{\rm turb}\sim L_{\rm edd}v_{\rm SMNR,rs}/3$ is the diffusion coefficient due to the turbulence around the reverse shock. 
If the eddy size is $L_{\rm edd}\sim 0.1\,{\rm pc}$, the injection fraction becomes $l_{\rm turb}/\Delta R \sim 0.5$. 
Although we have to understand turbulence in the SNR to estimate the injection fraction precisely, 
about $10-50\%$ of particles accelerated at the forward shock of the SNR could be transported to the reverse shock by the turbulent diffusion. 

The turbulent diffusion around the reverse shocked region can also move particles outward from the reverse shock front.    
In fact, particles that stay in the reverse shocked region can move to the forward shock region again by the turbulent diffusion. 
However, once particles escape into the freely expanding region and go back to the reverse shocked 
region, the particles can go back to the freely expanding region again in the timescale of $D_{\rm B}/cu_2$. 
The residence timescale in the downstream region of accelerating particle, $D_{\rm B}/cu_2\sim r_{\rm g}/u_2$, is typically much smaller than the eddy turnover time, $\sim L_{\rm edd}/u_2$, where $r_{\rm g}$ is the gyroradius. 
Therefore, once particles diffuse to the freely expanding ejecta, their turbulent diffusion can be neglected, that is, our Monte Carlo approach in section 2 is valid for particles that have already injected at the reverse shock front.  

\section{Discussion} 
We first discuss on the energy source of our model. 
The required energy per SNR is about $10^{50}$\,erg in order to supply Galactic CRs with an energy of $1$\,GeV. 
Since the recent CR observations show that the source spectrum of Galactic CRs should be $dN/dE \propto E^{-2.4}$, 
the required energy to supply PeV CRs is about $4 \times10^{47}$\,erg. 
In our model, the main acceleration is due to the PWN compression by the SNR. 
The energy source is the work done by the SNR, which is given by 
\begin{equation}
pdV \sim 10^{50}\,{\rm erg} \left( \frac{E_{\rm SN}}{10^{51}\,{\rm erg}}\right) \left( \frac{R_{\rm PWN}(t_{\rm c}) / R_{\rm SNR,fs}}{0.5} \right)^3~~, 
\end{equation}
where $p \sim E_{\rm SN} / R_{\rm SNR,fs}^3$ and $dV \sim R_{\rm PWN}(t_{\rm c})^3$ are the SNR pressure and the PWN volume compressed by the SNR. 
Therefore, the PWN-SNR system has enough energy to supply the PeV CRs.

We next discuss on the acceleration of heavy nuclei. 
CRs are organized not only by protons but also by heavy nuclei whose origin is also a long standing problem \citep{2016PhRvD..93h3001O}. 
Since the supernova ejecta is metal rich, 
the reverse shock propagating into the supernova ejecta is thought to be the origin of heavy CR nuclei \citep{2013ApJ...763...47P}. 
However, the maximum energy of the accelerated particle at the reverse shock is not so large because 
the magnetic field in the expanding supernova ejecta is expected to be very small. 
Furthermore, particles accelerated by the reverse shock suffer the adiabatic cooling.  
Our reacceleration model can boost the maximum energy of accelerated heavy nuclei, 
so that the PWN-SNR system could be important for the production of heavy CR nuclei.

Next, we discuss the energy spectrum of accelerated particles. 
In this paper, to investigate whether the PWN-SNR system can accelerate CRs to the PeV scale or not, we considered only the impulsive injection of CRs with $E=0.1$\,PeV at $t=10^3$\,yr. 
In reality, CRs accelerated by the SNR would be continuously injected with an energy spectrum to the PWN. 
Furthermore, since there is the potential difference of about $1$\,PV in the PWN, 
protons could be accelerated to $1$\,PeV by drifting the toroidal magnetic field \citep{1992MNRAS.257..493B,1996MNRAS.283.1083B}, 
which was not considered in this paper. 
Hence, further studies are needed to understand the energy spectrum of particles accelerated by the PWN-SNR system.

In addition, to understand the source spectrum of Galactic CRs that are injected to our Galaxy, we have to consider 
the spectrum of particles that have escaped from the SNR \citep{2010A&A...513A..17O,2011ApJ...729L..13O}, and variety of the PWN-SNR system. 
The PWN could expand again after the end of the PWN compression. 
We did not solve the particle transport and the energy loss during the later expansion phase in this paper. 
The time at which the PWN completes the re-expansion is about twice of the epoch when the PWN size 
becomes minimum \citep{2009ApJ...703.2051G}. 
Therefore, CRs inside the PWN-SNR system may lose $1- (1/2)^{2/5} \sim 24 \%$ of their energy due to the 
re-expansion of the PWN and the expansion of the SNR. 
When and how accelerated particles escape from the PWN-SNR system are important issues. 
These issues could be addressed by gamma-ray observations of the PWN-SNR system like SNR G327.1-1.1, W44 and so on.

In this paper we assumed as a first step, spherical symmetry for the PWN-SNR system and 
the Bohm scattering with constant magnetic field strength for the random walk. 
In reality, the magnetic field strength is not constant, a pulsar has a kick velocity, and the supernova ejecta, the ambient matter, the PWN have asymmetry.
In addition, the Rayleigh-Taylor instability amplifies the asymmetry \citep{2004A&A...420..937V}, 
so that the PWN-SNR system is actually more complicated. 
In particular, the strong turbulence could play important roles, that amplifies the magnetic field, 
affecting the particle motion \citep{2016MNRAS.460.4135P}, and accelerates particles by turbulent acceleration \citep{2013ApJ...767L..16O}. 
The turbulence eventually decays, so that accelerated particles can escape from the PWN.
On the other hand, the magnetic field is compressed by the PWN compression. Then, the escape time scale due to diffusion becomes long, so that more particles are accelerated by the PWN compression. 
In order to address above problems, we need a more realistic magnetohydrodynamical simulation.  
This will be addressed in future work.  
\section{Summary} 
We have proposed that PWNe inside SNRs are the CR PeVatron. 
Firstly, the SNR shock accelerates protons to $\sim 0.1$\,PeV. 
Then, the protons diffuse into the interior of the SNR and are reaccelerated to $\sim 0.2$\,PeV by the back and forth motion between the SNR and the PWN. 
Finally, the protons diffuse into the PWN and are accelerated to $\sim 1$\,PeV by the adiabatic compression while the PWN is compressed by the SNR. 
Our model predicts that there must be some structures in the spectrum of CR protons around $0.1$\,PeV. 
In addition, we have argued that the PWN-SNR system could be the origin of heavy CR nuclei.    
\section*{Acknowledgements}
Numerical computations were carried out on the XC30 system at the Center for Computational Astrophysics (CfCA) of the National Astronomical Observatory of Japan. 
This work was supported by JSPS KAKENHI Grant Number JP16K17702(YO), JP16J06773(SK), JP15K05088(RY), and JP18H01232(RY). Y.O. is supported by MEXT/JSPS Leading Initiative for Excellent Young Researchers. 



\bibliographystyle{mnras}
\bibliography{ref.bib} 


\bsp	
\label{lastpage}
\end{document}